\documentclass[a4paper,oneside,12pt]{elsarticle}
\usepackage{amsthm}
\usepackage{amsmath}
\usepackage{graphicx}%
\usepackage{amsfonts}%
\usepackage{amssymb}
 \usepackage{verbatim}
 \usepackage{float}
\usepackage{rotating}
\usepackage{amsmath,amsthm,amssymb}

\begin{document}
\title{\huge{\textbf{Optimisation of Stochastic Programming by Hidden Markov Modelling based Scenario Generation}}\\
\large{\textbf{by\\
Sovan Mitra}}} \maketitle
\section*{\large{Abstract}}
This paper formed part of a preliminary research report for a risk
consultancy and academic research. Stochastic Programming models
provide a powerful paradigm for decision making under uncertainty.
In these models the uncertainties  are  represented by a discrete
scenario tree and the quality of the solutions obtained is
governed by the quality of the scenarios generated. We propose a
new technique to generate scenarios based on Gaussian Mixture
Hidden Markov Modelling. We show that our approach  explicitly
captures important time varying dynamics of stochastic processes
(such as autoregression and jumps) as well as non-Gaussian
distribution characteristics (such as skewness and kurtosis). Our
scenario generation method enables richer robustness and scenario
analysis through exploiting the tractable properties of Markov
models and Gaussian mixture distributions. We demonstrate the
benefits of our scenario generation method by conducting numerical
experiments on FTSE-100 data.
\\\\
\textbf{Key words}: Markov Processes, Risk Analysis, Stochastic
Programming, Scenarios.

\line(1,0){400}

\section{Introduction}
Businesses are constantly confronted with the position of making
decisions under uncertainty e.g. portfolio investment, power
production and even agricultural planning \cite{birge1997isp}.
Therefore the ability to be able to quantify and optimise such
decisions is of paramount value and the modelling paradigm of
Stochastic Programming offers such a possibility. Stochastic
Programming (SP)  has been proven to provide practical and optimal
solutions e.g. Mulvey's celebrated Towers Perrin-Tillinghast model
saved US West \$450-\$1000 million \cite{MULVAss}.
\par
Stochastic Programming concerns decision making under uncertainty.
A multistage stochastic programming problem is defined as:
\begin{eqnarray}
\min_{x \in \Xi} \mathbb{E}_{P}\{f(x,\omega)\} = \min_{x \in \Xi}
\int_{\Omega} f(x,\omega)dP(\omega),
\end{eqnarray}
where
\begin{itemize}
  \item x=$\{x_{t}\}$ is the set of decisions at each stage
  t=1,2....T
  where T is the total number of stages;
  \item $\Xi$ is the set of feasible decisions;
  \item $\omega$ is the stochastic process;
  \item $\Omega$ is the set of all possible events;
  \item P is the probability measure for stochastic process
  $\omega$;
    \item $\mathbb{E}_{P}$ is the expectation taken with respect to
    probability measure P;
    \item $f(x,\omega)$ is the cost function.
\end{itemize}
In stochastic programming a method is required to discretise the
random process $\omega$ into a set of discrete outcomes, known as
scenarios. The method of obtaining scenarios is known as scenario
generation.
\par
The root node of the scenario tree represents `today' and is known
with certainty. Nodes or scenarios at later stages progressing
from the root node represent possible realisations of the random
process, where each scenario has a probability and a value. As
Birge and Louveaux \cite{birge1997isp} point out, scenarios are
particularly useful when the optimal solution varies considerably
with changes in value of the stochastic variables. If this were
not the case, the solution obtained for one stochastic variable
realisation ought to be a good solution for most realisations.
\par
The general formulation of equation (1) under scenario generation
is:
\begin{eqnarray*}
\min_{x \in \Xi} \mathbb{E}_{P}\{f(x,\omega)\} &=& \min_{x \in
\Xi} \sum_{i=1}^{i=N_{1}} p^{i}_{1} f(x_{1}^{i},\omega_{1}^{i})+
\sum_{i=1}^{i=N_{2}}p^{i}_{2}f(x_{2}^{i},\omega_{2}^{i})+..........\\
 &+& \sum_{i=1}^{i=N_{T}} p^{i}_{T}f(x_{T}^{i},\omega_{T}^{i}),
\end{eqnarray*}

where
\begin{itemize}
\item \textit{i=1,2....$N_{t}$} is the scenario and \textit{$N_{t}$} is the total number of
scenarios at stage \textit{t};
\item $p^{i}_{t}$ is probability of scenario \textit{i} at stage
\textit{t};
\item $f(x^{i}_{t},\omega^{i}_{t})$ is the cost function of scenario \textit{i} at
stage \textit{t}.
\end{itemize}

A key issue in scenario generation that tends not to be addressed
is the ability to capture time varying dynamics of the random
process $\omega$ (such as jumps and reversions) and to a lesser
extent, capturing non-Gaussian characteristics. Our scenario
generation method specifically addresses the aforementioned
problems through the properties of Gaussian Mixtures (GM) and
Hidden Markov Models (HMM).
\par
In GM HMM we model a time series as a HMM where each hidden Markov
state is represented by a univariate mixture of Gaussians. To
produce a scenario tree we let each Markov state correspond to a
stage in the scenario tree. Then we obtain scenarios at each
scenario tree stage by sampling from each state's associated
univariate mixture of Gaussians. Hence the key part in our
scenario generation procedure is determining the GM HMM and then
selecting a state sequence for our scenario tree. Once the states
have been selected, we simply sample from each state's GM to
obtain scenarios at each stage.
\\\\
\emph{Guided Tour}\\
The outline of this paper is as follows: firstly we review current
scenario generation methods and then introduce the GM HMM model.
We then explain the advantages of the GM HMM for scenario
generation, showing how the particular features of GM and HMM
capture the time varying dynamics of data. We also show how GM HMM
provides SP modelling benefits e.g. robustness analysis. Finally,
we illustrate the GM HMM scenario generation and modelling
benefits by applying it to the FTSE-100 index, giving
computational results.

\section{Current Scenario Generation Methods}
Currently there exist four  main categories for scenario
generation \cite{didomenica:spa}: sampling, simulation,
statistical property matching and hybrid methods e.g.
simulation-optimisation \cite{april2003sbo}. For a detailed list
of scenario generation methods see
\cite{didomenica:spa},\cite{MitraWPOptiSG}.

In simulation, a stochastic process's simulation results in a
``path by path'' production of the scenario tree \cite{kaut:esg}.
To reduce the number of paths sometimes paths are bounded together
by some grouping method, for instance \cite{gulpinar2004sao}.
Popular stochastic processes include geometric Brownian motion and
its variants.

In scenario generation by sampling we sample values from a
distribution and the sampled value represents the scenario's
value. Various methods of sampling exist \cite{april2003sbo},
giving rise to different sampling scenario generation methods.
Popular methods include: random sampling, importance sampling
\cite{dempster1999ebi}, bootstrap sampling, internal sampling
\cite{kaut:esg},conditional sampling and stratified sampling.
\par
In scenario generation by statistical property matching we do not
require knowledge of the random variable's ($\omega$) pdf
(probability density function). Instead we require the statistical
properties of the generated scenario tree to match the statistical
properties of some target (as closely as possible) e.g.
percentiles. Hoyland and Wallace in \cite{hoyland2001gst} describe
a scenario generation method using moment matching.

\section{Gaussian Mixture Hidden Markov Model Scenario Generation}
\subsection{Markov Models}
Markov Models (MM)   and Hidden Markov Models (HMM) are methods of
mathematically modelling time varying dynamics of some statistical
process. They only require a weak set of assumptions yet provide
powerful results. MM and HMM model a stochastic process (or any
system) as a set of states with each state possessing a set of
signals or observations. Changes in the stochastic process or
system are modelled by movements between states. The models have
been used in  diverse  applications such as  Finance
\cite{hardy2001rsm}, queuing theory \cite{salih2006maa},
Engineering \cite{trentin2001sha} and biological modelling
\cite{melodelima2006cpi}.

A Markov model (or more specifically a Markov chain) is a
stochastic process $X_{t}$ with a countable set of states and
possesses the Markov property. That is, given $X_{t}$ is in state
$i_{t}$ (denoted $X_{t}=i_{t}$), $X_{t+1}$ is not influenced by
the set of values $X_{u}$ for $u<t$. Formally, the Markov property
is
\[
P(X_{t+1}=j\mid
X_{0}=i_{0},X_{1}=i_{1},...,X_{t}=i_{t})=P(X_{t+1}=j\mid
X_{t}=i_{t}),
\]
where $i_{t}$ denotes a state at time t. As time passes the
process may remain or change to another state (known as state
transition). The state transition probability matrix (also known
as the \emph{transition kernel}) tells us these probabilities
$P_{ij}=P(X_{n+1}=j \mid X_{n}=i)$ where $P_{ij}$ is the
probability of the process changing to state \emph{j} given that
we are now in state \emph{i}. Note that we assume all
probabilities are stationary in time.

\subsection{Gaussian Mixture Hidden Markov Models}
In Markov models we directly observe the state associated with
each signal or observation; whereas  in Hidden Markov models  the
states are directly unidentifiable or unobservable. However in HMM
we assume that each signal or observation is a probabilistic
function of each hidden state, hence we can statistically infer
the HMM states from observations. If observations were not
statistical functions of state then we could not determine the
hidden states. The observation probabilities are represented in
the observation matrix B, showing each observation probability for
each state. Note that as in plain Markov models we assume all
probabilities are stationary.
\par
A hidden Markov model is fully defined when we know
\cite{rabiner1989thm}:
\begin{itemize}
  \item \emph{N} is the  total number of hidden states
  $S_{1},S_{2}...,S_{N}$;
\item \emph{A} represents the hidden state transition matrix of size NxN. Each
element is $a_{ij}=P(q_{t+1}=S_{j}|q_{t}=S_{i})$ where $q_{t}$ is
the hidden state at time t;
  \item $\pi_{i}$ (=$P(q_{0}=S_{i}$)) for $1 \leq i \leq N $ denotes the
  the probability of being in state $S_{i}$ at time \emph{t=0};
\item \emph{C} denotes the number of distinct  observations. The
set of  observations  are represented as $v_{1},v_{2} \ldots
,v_{C}$. For example  if we model weather as a HMM with only 2
observation signals hot and cold, then \emph{C} = 2 and
$v_{1}$=hot, $v_{2}$=cold;
\item
\emph{B} denotes the observation matrix. Each entry is
$b_{j}(c)=P(v_{c}|q_{t}=S_{j})$, $1 \leq c \leq C$. So $b_{j}(c)$
is the probability of observation $v_{c}$ at time \emph{t}, given
that we are in state $S_{j}$ at time \emph{t}.
%The emission probs may be given but if not then they must be
%calculated/inferred.
%Note that a signal emission that causes no state transition simply
%means the arc for that signal transmission "re-loops" back to the same state.
\end{itemize}
%In our model, we assume observations v are modelled by a
%univariate mixture of Gaussians.
A Gaussian mixture is a pdf composed of a weighted sum of
univariate Gaussians, where all the weights sum to one. Let
$\xi(\omega)$ be a random variable $\xi(\omega) \in \mathbb{R}$
whose pdf $f(\xi)$ is represented by \emph{m} Gaussian mixtures \cite{everitt1981fmd}: %$f(\xi)$=f(x) effectively
\begin{eqnarray}
f(\xi)=w_{1}N(\xi,\mu_{1}, \sigma_{1})+ w_{2}N(\xi,\mu_{2},
\sigma_{2}) +....+ w_{m}N(\xi,\mu_{m}, \sigma_{m}),
\end{eqnarray}
where $\mu_{m},\sigma_{m},w_{m}$ denotes the \emph{m}th Gaussian's
mean, standard deviation and weighting respectively.
%\begin{itemize}
%  \item $f(x)$=pdf of v
%  \item $c_{k}$ mixture weighting for Gaussian mixture k, where
%  $\sum^{m}_{k=1}c_{k}$=1, $c_{k} > 0  \forall k$
%  \item $\mu_{k}$=mean for each mixture k
%\item $\sigma_{k}$ standard deviation for each Gaussian k where
%$1\leq k \leq m$
%\item m is total number of Gaussian mixtures
%\end{itemize}
Note that the  mean and all other moments are weighted sums too
\cite{everitt1981fmd},
for instance the overall distribution's mean $\mu_{f}$ is: %NOTE: variance NOT moment but moment about mean
\begin{eqnarray}
\mu_{f}&=&\sum_{k=1}^{m} w_{k}\mu_{k}.
\end{eqnarray}
%Var[x]&=&\sum_{k=1}^{m}w_{k}Var_{k}(x)+[\sum_{k=1}^{m}E_{k}(x)^{2}-(\sum_{k=1}^{m}E_{k}(x)^{2})^{2}]
%\cite{MM}
Now a Gaussian mixture HMM is simply a HMM where each state's
observations is represented by a Gaussian mixture pdf. In other
words $f_{j}(\xi)$ is our observation signal $b_{j}(c)$  where
\cite{levinson2005mms}:
\begin{eqnarray}
f_{j}(\xi)=\sum_{k=1}^{m} w_{jk}N(\xi,\mu_{jk}, \sigma_{jk}), \xi
\in \mathbb{R},
\end{eqnarray}
where
\begin{itemize}
  \item $f_{j}(\xi)$ is the pdf of $\xi$ in state j where $1\leq j \leq
  N$;
  \item $w_{jk}$ is the mixture weighting in state j for Gaussian mixture
  k where $1\leq k \leq m$, $w_{jk} \geq 0$ $\forall j,k$ and  $\sum^{m}_{k=1}w_{jk}$=1 $\forall
  j$;
  \item $\mu_{jk}$ is the mean for each mixture \emph{k}, in state
  \emph{j};
\item $\sigma_{jk}$ is the  standard deviation  for each Gaussian
\emph{k} in state \emph{j};
\item \emph{m} is the total number of Gaussians in the mixture.
%\item n is total
%number of states
\end{itemize}

\subsection{GM HMM Scenario Generation Method}
Given we are applying GM HMM to financial data we may expect the
stock market to belong to 3 economic phases (growth, recession and
a transitional state) so we set N=3. Additionally, we choose
\emph{m}=2 given that most empirical financial return
distributions can be described by a mixture of 2 Gaussians. We
should also note by the Akaike information criterion that it is
better to minimise the number of variables to prevent overfitting.
\\\\
\textit{Scenario Generation Algorithm}\\
Let $\hat{r}$ where $\hat{r}\sim U[0,1]$ denote a random number
drawn from a unit interval.
%%%%%%%%%%%%%%%%%%%%%%%%%%%%%%%%%%%%%%%%%%%%%%%%%%%%%%%%%%%%%%%%%%%%%%%%%%%
\begin{comment}
\begin{itemize}
\item $\hat{r}$ where $\hat{r}\sim U[0,1]$ denote a random number
drawn from a unit interval;
\item $j \in \{1,2,3\}$ denote the HMM state;
\item $a_{j1}$ is the
state transition probability to state 1 given we are currently in
state j $(a_{j1}=p(q_{t+1}=1|q_{t}=j))$.
\end{itemize}
\end{comment}
%%%%%%%%%%%%%%%%%%%%%%%%%%%%%%%%%%%%%%%%%%%%%%%%%%%%%%%%%%%%%%%%%%%%%%%%%%%

\begin{enumerate}

\item Randomly select an initial state:\\
Draw a random number $\hat{r}$.\\
If:
\begin{itemize}
\item $\hat{r}<\pi_{1}$ set  j=1;

\item $\pi_{1}< \hat{r}<\pi_{1}+\pi_{2}$ set  j=2;

\item otherwise set j=3.
\end{itemize}

\item Randomly select a Gaussian component k (\cite{everitt1981fmd},\cite{bignami1971nsc}):\\
Draw a random number $\hat{r}$. \\
Set k=1 if $\hat{r}<w_{j1}$, otherwise k=2.

\item Sample Gaussian $N({\mu}_{jk},{\sigma}_{jk})$.

\item Randomly select the next state:\\
Draw a random number $\hat{r}$. \\
If:
\begin{itemize}
\item $\hat{r}<a_{j1}$ set  j'=1;

\item $a_{j1}< \hat{r}<a_{j1}+a_{j2}$ set  j'=2;

\item otherwise set j'=3.
\end{itemize}
Set current state to new state: $j \mapsto j'$.

\item Increment I.\\
If I equals the maximum number of scenarios required then stop,
otherwise goto step 2.

\end{enumerate}

To calibrate the GM HMM we apply the Baum-Welch algorithm, which
is an expectation-maximisation algorithm. The algorithm determines
the GM HMM model parameters such that it maximises the statistical
likelihood of a given observation data. The algorithm is
guaranteed to converge to some local optimum for calculating the
likelihood of a given observation sequence. The reader is referred
to \cite{rabiner1989thm},\cite{juang1991hmm},\cite{juang1985pdm}
for more information.

If we define $\lambda$=\{A, B, $\pi_{i}$\} to denote the set of
HMM parameters we wish to estimate then the pseudo code for the
Baum-Welch algorithm is:
\begin{enumerate}
\item Initialisation:\\
Input initial values of $\lambda$ (otherwise randomly initialise)
and
calculate $P(O|\lambda)$ using the forward algorithm. % $p(O|\lambda)$ relates to Forward NOT viterbi
\item Estimate new values of $\lambda$.\\
Iterate until convergence:
\begin{enumerate}
\item Using current $\lambda$ calculate $\gamma_{t}(i,j)=P(q_{t}=S_{i},q_{t+1}=S_{j}|O,\lambda)$ for each
\textit{t,i,j} where
\begin{eqnarray}
\gamma_{t}(i,j)=\dfrac
{\alpha_{t}(i)a_{ij}b_{j}(O_{t+1})\beta_{t+1}(j)}{\sum_{i=1}^{N}
\sum_{j=1}^{N}\alpha_{t}(i)a_{ij}b_{j}(O_{t+1})\beta_{t+1}(j)}.
\end{eqnarray}
Variables  $\alpha_{t}(i),\beta_{t}(j)$ are known as the forward
and backward variables respectively (see \cite{rabiner1989thm}):
\begin{eqnarray}
\alpha_{t}(i) &=& P(O_{1}O_{2}.....O_{t},q_{t}=S_{i}|\lambda),\\
\beta_{t}(j) &=& P(O_{t+1}O_{t+2}.....O_{T}|q_{t}=S_{j},\lambda).
\end{eqnarray}

\item Calculate  new $\lambda$ parameter estimates using
$\gamma_{t}(i,j)$ (see \cite{rabiner1989thm}). %,p.265 equations 40a-c,p.267 equations 52-4)\\
\item Calculate $P(O|\lambda)$ with new $\lambda$ values using the
forward algorithm. %-cf p265 text after eqn 40c \cite{rabiner1989thm}
\end{enumerate}
\item Goto step 4 if two consecutive calculations of
$P(O|\lambda)$ are equal (or converge within a specified range).
Otherwise repeat iterations: goto step 2.
\item Output $\lambda$.
 %cf p265 bottom.
%Recall system outlook: I/P: $\lambda$, control: algorithm, O/P:new  $\lambda$).
%\textbf{BW extension to continuous Multivariate Gaussian Mixture Distributions}\\
\end{enumerate}

\subsection{Scenario Generation Salient Features}

\subsubsection*{Explicit Modelling of Reversionary and Jump Dynamics}
Stochastic processes or time series tend to exhibit reversionary
characteristics (e.g. autoregression) and jumps. These dynamics
make continuous time modelling challenging and are typically
ignored in scenario generation since most methods are unable to
accommodate these dynamics. Using HMM we can explicitly capture
reversionary characteristics and jumps; in fact HMM have been used
widely in Finance to capture such aspects
\cite{rebonato2004vac},\cite{hardy2001rsm},\cite{hamilton1989nae}.
Important reversion properties captured are:

\begin{itemize}
\item autoregression:  this is captured by the transition probabilities of
returning to the same current Markov state $P(q_{t+1}=S_{i}|q_{t}=S_{i})$  (\cite{hamilton1989nae},\cite{hamilton2005rss}%p1-2
).
\item conditional
heteroscedasticity:% (variance changing on condition of the past):
since each Markov state has a different variance and each future
state's transition depends on the current state, variance changing
is conditional on the past.
\item cyclical behaviour \cite{hamilton1989nae}: the cyclical nature of data is captured
through the transition probability matrix A, so that the MM cycles
through its entire set of states.
\end{itemize}
Jumps are explicitly modelled by the transition probability of
switching between 2 states with significantly different distributions \cite{rebonato2004vac}.%p.798

\subsubsection*{Modelling Diverse Distributional Characteristics}
Gaussian mixtures can capture virtually any empirical
distribution. In fact GM have been used to accurately capture
outliers and ''noisiness" in empirical data
\cite{titterington1985saf}. Various distribution characteristics
are captured by simply adjusting the mixture component means,
variances and weighting.

\subsubsection*{Tractable Statistical Calculations}
Gaussian distributions tend to be applied to data modelling due to
their analytical tractability yet most (empirical) distributions
tend to be non-Gaussian e.g. asymmetric and fat tailed.
Additionally non-Gaussian models tend to possess intractable
statistical calculations e.g. the Levy distribution.
\par
Using a mixture of Gaussians we can model most non-Gaussian
distributional characteristics (e.g. skewness, kurtosis etc....)
without losing tractability. For example, the cumulative
distribution function for a GM mixture consisting of 0 mean
components is simply a weighted sum of each component's cumulative
distribution value. Thus the the overall cumulative distribution
function for any univariate GM $F(\xi)$ gives \cite{BuckPortOpt}:
\begin{eqnarray}
F(\xi)=\sum_{k=1}^{m}w_{k}\Phi\left(\dfrac{\xi-\mu_{k}}{\sigma_{k}}\right),
\end{eqnarray}
where $\Phi(z)$ is the \emph{standard normal} cumulative
distribution function for random variable z.

\subsubsection*{Copula Approximation} Copulas map marginal
distributions to 1 multivariate distribution and so are
particularly useful in  risk analysis. However copulas can become
analytically intractable, causing calculation and computation
difficulties e.g. sampling and scenario generation
\cite{hoyland2003hmm}. Furthermore, there exist significant
implementation problems since we would need to 1)estimate the
covariance matrix, which tends to be unstable over time 2)specify
the copula type (e.g. median, Gaussian) 3)determine a feasible
method of scenario generation from the overall multivariate
distribution. All these tasks are non-trivial.
\par
Mixture of Gaussians can be used as an alternative to copulas
\cite{BuckPortOpt}, where all the component distributions are
mapped into one univariate Gaussian mixture. Using univariate GM
the problematic steps of 1-2 are removed, rather only the
weighting for each component distribution need be estimated.
\par
If we map to a univariate (rather than a multivariate)
distribution then the distribution model is more amenable to
analytical calculations, computational methods and scenario
generation. For instance GM can be applied to models of geometric
Brownian motion \cite{brigo2003aap} whereas for multivariate
Gaussians this becomes non-trivial. Univariate GM can be applied
to a wider and more efficient range of scenario generation and
computational methods.

\subsubsection*{Robustness and Scenario Analysis}
In SP we may wish to evaluate a SP model's robustness by stress
testing and shock analyses \cite{birge1997isp}. Dupacova suggests
a contamination method for stress testing \cite{dupacova1996sbs}
where randomness is added to the scenarios. For shock analysis  we
typically expose the SP model to an extreme event or a \emph{worst
case scenario}.
\par
Using GM we can enrich stress and shock analysis, in fact GM
already have been used as robustness methods in statistical
modelling \cite{titterington1985saf} but not in SP. For stress
testing, we can add a GM of any choice to the overall
distribution, representing a contamination as in Dupacova's
method. Furthermore, we can calculate the impact on the overall
distribution's moments and pdf for given weights, means and
variances.
\par
For shock analyses, rather than simply testing with respect to an
extreme event scenario we can model a shock by a GM of chosen
weighting, mean and variance and this is more realistic. For
instance an overall GM distribution may represent a company's
profit and we may wish to determine the impact of  some extreme
event e.g. hurricanes can severely disrupt natural gas supplies.
We could more realistically model a hurricane event by a GM
compared to testing with respect to one extreme event scenario.

\subsubsection*{Adaptability to Multivariate Distributions}
The GM HMM can model multivariate Gaussian mixtures (see
\cite{levinson2005mms} for a detailed application in Engineering)
and there exist significant statistical literature, algorithms and
programs available. Additionally, multivariate GM HMM capture time
varying covariance matrices by specifying one for each hidden
Markov state \cite{rabiner1989thm}. This is a significant
advantage over other multivariate modelling methods particularly
for financial applications, where correlations are critical to
portfolio diversification.

\section{Numerical Experiment: Case Study Application to the FTSE-100}
We present results of applying the GM HMM to modelling the monthly
returns of the FTSE-100 $r_{t}$,  where $r_{t}$ is defined by
\cite{HULLIntro}:
\begin{eqnarray}
r_{t}=ln\dfrac{I_{t}}{I_{t-1}}.
\end{eqnarray}
The variable $I_{t}$ is defined by the FTSE-100 index value in
month t. We used monthly data from April 1984-February 2007; the
GM HMM was calibrated by the Baum-Welch algorithm
\cite{rabiner1989thm} using an implementation by Murphy
\cite{MurphyHMMProgram}. The results are now presented.

\pagebreak[4]
%\section{Computational Results}
\begin{table}[h!b!p!]\caption{ Transition Matrix A Results}
\begin{center}
\begin{tabular}{|c|c|c|c|}
\hline
     &  \multicolumn{3}{|c|}{Next State}\\\cline{2-4}
Current State & State 1 & State 2 & State 3 \\\hline
  State 1  & 0.71 & 0.057 & 0.23 \\
  State 2  & 0.13 & 0.51 & 0.36 \\
  State 3  & 0.45 & 0.19 & 0.36 \\ \hline
\end{tabular}
\end{center}
\end{table}

\begin{table}[h!b!p!]\caption{Mixture Means $\mu_{jk}$ (monthly \%
return)}
\begin{center}
\begin{tabular}{|c|c|c|c|}
\hline
  Component Distribution (k) & State 1 & State 2 & State 3 \\\hline
   1 & 1.662  & -2.588 & 0.776 \\
   2 & -30.170 & 1.113 & -0.680 \\
  Overall  & 1.41   & 0.097 & 0.44\\ \hline
\end{tabular}
\end{center}
\end{table}

\begin{table}[h!b!p!]\caption{Standard Deviations $\sigma_{jk}$ (monthly \%
return)}
\begin{center}
\begin{tabular}{|c|c|c|c|}
\hline
  Component Distribution (k) & State 1 & State 2 & State 3 \\\hline
   Gaussian 1 & 1.201  & 2.940 & 2.447 \\
  Gaussian 2 & 0.100 & 2.728 & 2.715\\
  Overall   & 1.192 & 2.88 & 2.51\\ \hline
\end{tabular}
\end{center}
\end{table}

\begin{table}[h!b!p!]\caption{Initial State Probabilities ($\pi_{i}$)}
\begin{center}
\begin{tabular}{|c|c|}\hline
  State & Probability \\\hline
  1 & 1.0434e-024\\
  2 & 0.99\\
  3 & 1.6637e-017\\ \hline
\end{tabular}
\end{center}
\end{table}

\begin{table}[h!b!p!]\caption{Weighting Matrix ($w_{jk}$)}
\begin{center}
\begin{tabular}{|c|c|c|}\hline
  State & Component k=1 & Component k=2 \\\hline
  1 & 0.99207 & 0.0079285 \\
  2 & 0.73889 & 0.26111 \\
  3 & 0.7665 & 0.2335 \\ \hline
\end{tabular}
\end{center}
\end{table}
\newpage

\begin{table}[h!b!p!]\caption{Stress Test Example Scenario Tree}
\begin{center}
\begin{tabular}{|c|c|c|}\hline
Scenario Number & Unstressed & Stressed\\\hline
1 & -0.2521  & -10.1146 \\
2 & -0.0686   & -10.0433\\
3 & 0.0572  & -0.3384\\
4 & 0.8312 & 1.6168\\
5 & 2.0156  & 1.8125\\
6 & 2.3479 & 1.8717\\
7 & 2.5199 &  2.0075\\
8 & 2.6925 &  2.0551\\
9 & 3.1681 &  3.0902\\
10 & 3.6119 &  3.0923\\ \hline
\end{tabular}
\end{center}
\end{table}

\begin{table}[h!b!p!]\caption{Scenario Generation Under Different Credit Risk
Weightings}
\begin{center}
\begin{tabular}{|c|c|c|c|}\hline
  % after \\: \hline or \cline{col1-col2} \cline{col3-col4} ...
Scenario Number & \multicolumn{3}{|c|}{Credit Risk
Weighting}\\\cline{2-4}
 & 0.1 & 0.3 & 0.5\\ \hline
1 & -0.3384 & -0.3384 & -0.3384 \\
2 & -0.1115 & -0.1115 & -0.1115\\
3 & -0.0967 & -0.1043 & -0.1043\\
4 & -0.0881 & -0.0987 & -0.0987\\
5 & -0.1043 & -0.0967 & -0.0983 \\
6 & 1.6168 & -0.0881 & -0.0971\\
7 & 1.8125 & 1.6168 & -0.0967 \\
8 & 1.8717 & 1.8717 & -0.0881\\
9 & 2.0075 & 2.0075 & -0.0881\\
10& 3.0902 & 3.0902 & 1.6168\\\hline
\end{tabular}
\end{center}
\end{table}

\newpage
Figure 1: FTSE-100 Empirical Monthly Returns (\%) for each Month
\footnote{from May 1984}\\
\includegraphics[height=80mm,width=150mm]{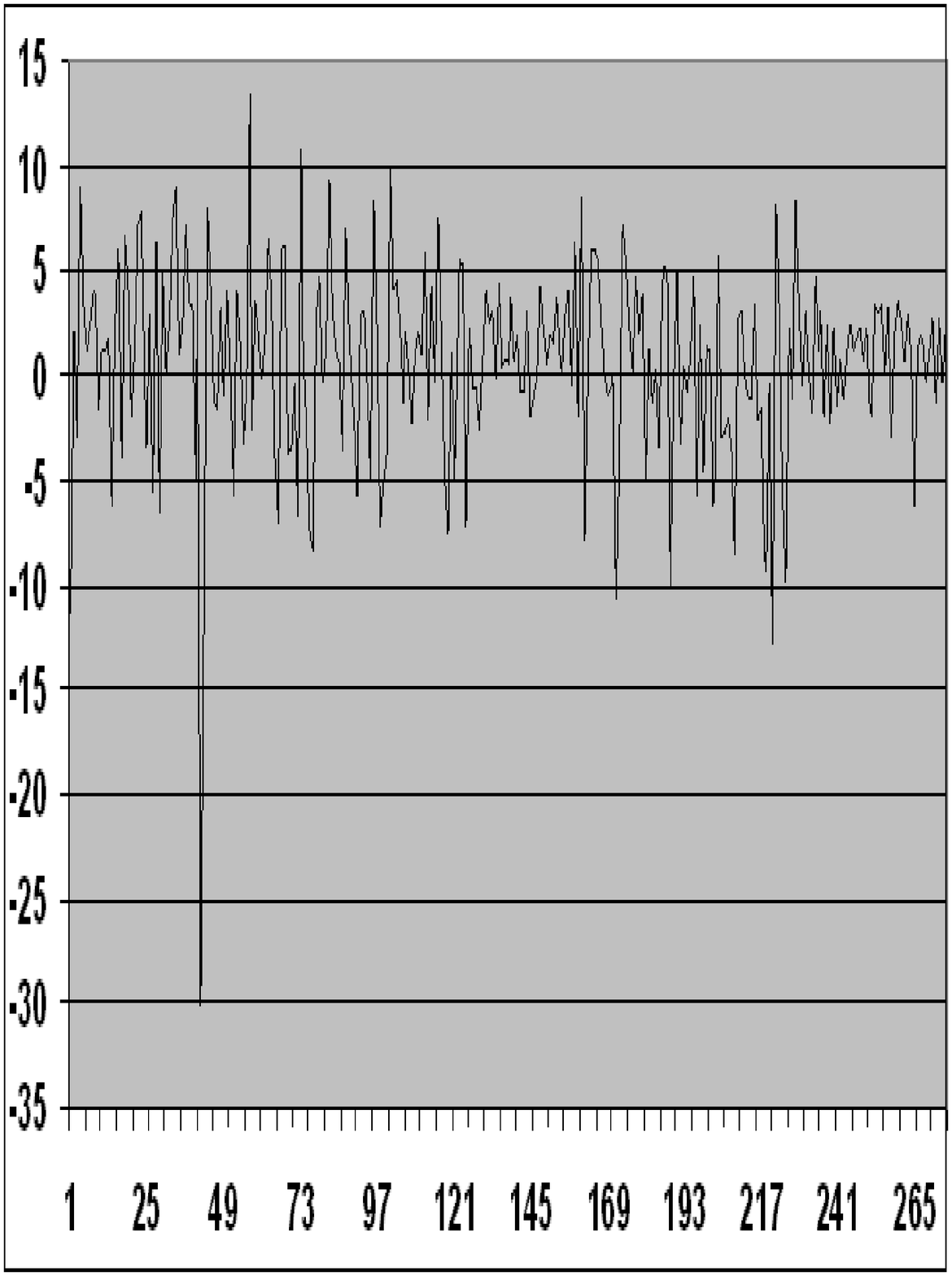}
\\\\\\
\\\\\\
Figure 2: GM HMM Simulation of FTSE-100 Monthly Returns (\%)\\ %  (y-axis: monthly return \% ,x-axis:time (months))\\
\includegraphics[height=80mm,width=150mm]{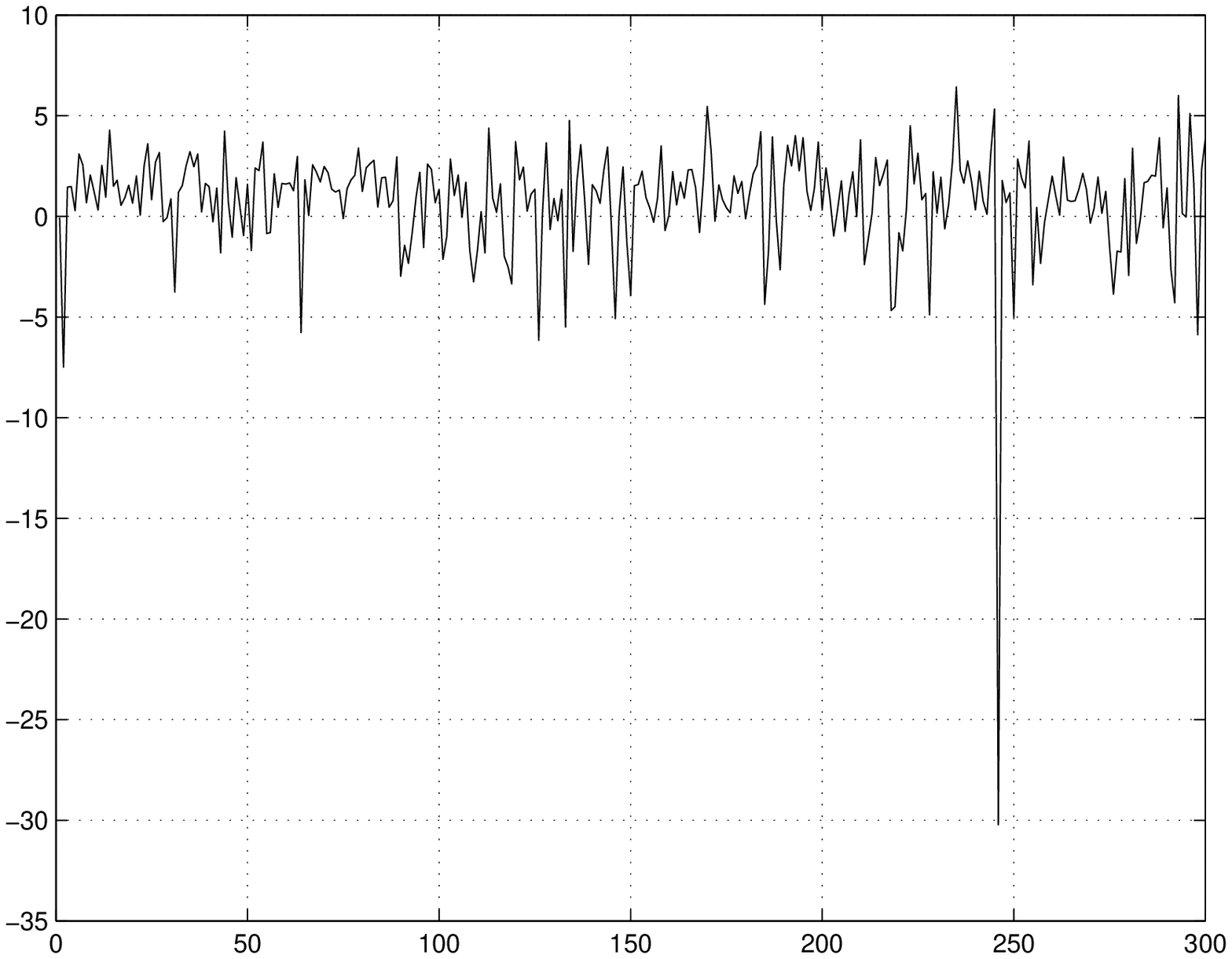}

\newpage
From table 2's results  we can see state 1 is the growth state,
state 2 the recession state, state 3 is transitional (since the
overall distribution means are positive, negative and about 0
respectively). We see from table 1 that if we are in state 2, then
51\% of the time we will remain in state 2, similarly 71\% for
state 1.
%Notice the state pdfs having a mixture of Gaussians enable us to capture important non-Gaussian characteristics for
%each state's distribution e.g. fat tails and asymmetries. Therefore when sampling each state's distribution we can obtain
%more representative scenarios, in particular worst case scenarios.
\par
We provide a GM HMM simulation of the FTSE-100 in figure 2 to
compare its modelling performance to the empirical data (see
figure 1). Simulating the GM HMM involves randomly selecting a
Markov state in accordance to the transition matrix A and at each
state randomly sampling a value from its pdf. Notice that the
model captures many key features of the empirical data, namely
extreme losses, autoregression and jumps.
\\\\
\textit{Copula Approximation}\\
Copulas are used  to see a factor's impact upon the overall
multivariate distribution e.g. mean, variance etc... . For
example, we may wish to determine the impact upon a FTSE-100 index
tracking portfolio (or any portfolio) from credit risk or
operational risks \cite{voit2003bmo}. Rather than applying copulas
to analyse the impact of operational and credit risks,  we can
apply a GM model and represent these two factors as two Gaussian
components. For example, if we choose to examine the impact on
state 1 then we may have the distribution:
\\\\
\begin{tabular}{|c|c|c|c|}\hline
  Risk Factor & Mean & Sigma & Weighting \\\hline
FTSE-100 Portfolio Return Component Gaussian 1 & 1.66 & 1.2 &  0.69\\
FTSE-100 Portfolio Return Component Gaussian 2 & -30 & 0.1 &   0.01\\
  Operational Risk & 0 & 1 & 0.2 \\
  Credit Risk & -0.1 & 0.01 & 0.1 \\ \hline
\end{tabular}
\\\\
We can also calculate the moments, the pdf is a simply weighted
sum:
\begin{eqnarray}
f(\xi)= 0.7N(\xi,1.41,1.2) + 0.2N(\xi,0,1) + 0.1N(\xi,-0.1,0.1).
\end{eqnarray}
Since the overall distribution is univariate, standard GM sampling
and scenario generation techniques can be applied. This
demonstates the simplicity and computational advantages of GM
models compared to copulas.
\\\\
\textit{Factor Specific Stress and Shock Testing (Robustness Analysis)}\\
Rather than conducting a worst case scenario analysis for shock
testing, we can more accurately model a ``shock event" (such as a
``credit crunch") with a Gaussian mixture distribution component
of chosen weighting, mean and variance. For instance we may
propose to stress test state 1's distribution using:
\begin{eqnarray}\label{shock eqn}
f(\xi)= 0.9(0.99N(\xi,1.66,1.2) + 0.01N(\xi,-30,0.1)) +
0.1N(\xi,-10,0.1),
\end{eqnarray}
where the first two Gaussians represent state 1's (unstressed)
distribution and the third represents an extreme event. Notice
that the extreme event's inclusion fattens the left tail of the
overall distribution so \emph{all} scenarios sampled from this
distribution are affected, unlike in standard worst case scenario
analysis. We show a scenario tree generated from equation
\eqref{shock eqn} in table 6; notice the presence of more negative
events scenarios due to the ``shock" event Gaussian.
\par
Using Gaussian mixtures, we can conduct robustness analysis with
respect to 1 particular risk factor and calculate the impact on
the overall distribution. For example we vary the credit risk's
weighting by 0.1,0.3,0.5 and show the impact on scenario
generation in table 7. As can be seen in table 7, as we increase
the credit weighting the impact of negative scenarios becomes more
prevalent.
%%%%%%%%%%%%%%%%%%%%%%%%%%%%%%%%%%%%%%%%%%%%%%%%%%%%%%%%%%%%%%%%%%%%%%%%%%%%%%%%%%%%%%%%%%%%%%%%%%%%%%%%%%%%%%%%%%%%%%%%%%%%%
\begin{comment}
\\\\
Table 9: Credit Risk Stress Testing Weightings
\begin{footnotesize}
\begin{eqnarray*}
\begin{tabular}{|c|c|c|c|c|}\hline
Gaussian 1 & Gaussian 2  & Operational Risk & Credit Risk\\\hline
 0.69 & 0.1  & 0.2 & 0.1\\
0.54  & 0.01  & 0.15 & 0.3\\
0.389 & 0.001 & 0.11 & 0.5\\\hline
\end{tabular}
\end{eqnarray*}
\end{footnotesize}
\end{comment}
%%%%%%%%%%%%%%%%%%%%%%%%%%%%%%%%%%%%%%%%%%%%%%%%%%%%%%%%%%%%%%%%%%%%%%%%%%%%%%%%%%%%%%%%%%%%%%%%%%%%%%%%%%%%%%%%%%%%%%%%%%%%%

\section{Conclusion}
In this paper, a new method of scenario generation has been
proposed and demonstrated. We have shown how the properties of GM
HMM enable us to capture important time varying dynamics of data,
in particular reversion and jumps. This enables us to produce more
effective scenario trees. The tractable properties of GM enable us
to further explore SP models, through richer stress and shock
testing with respect to particular factors. Additionally we offer
a simple method of approximating copulas for risk analysis. We
have demonstrated these properties of GM HMM scenario generation
by numerical experiments on FTSE-100 data.

\newpage
\bibliographystyle{plain}
\addcontentsline{toc}{section}{References}
\bibliography{Ref}

\end{document}